\documentclass[twocolumn,superscriptaddress,aps,preprintnumbers,amsmath,amssymb,prl,nofootinbib]{revtex4}

\usepackage{graphicx}
\usepackage{epstopdf}
\usepackage{dcolumn}
\usepackage{bm}
\usepackage{hyperref}
\usepackage{amsmath,amsfonts,amssymb}
\usepackage{color}

\begin{document}

\preprint{YITP-18-84, IPMU18-0129}
\title{
Blue-tilted Primordial Gravitational Waves
from Massive Gravity
}

\author{Tomohiro Fujita}
\affiliation{Department of Physics, Kyoto University, Kyoto, 606-8502, Japan}
\author{Sachiko Kuroyanagi}
\affiliation{Department of Physics, Nagoya University, Chikusa, Nagoya 464-8602, Japan}
\author{Shuntaro Mizuno}
\affiliation{Center for Gravitational Physics, Yukawa Institute for Theoretical Physics, Kyoto University, Kyoto 606-8502, Japan}
\author{Shinji Mukohyama}
\affiliation{Center for Gravitational Physics, Yukawa Institute for Theoretical Physics, Kyoto University, Kyoto 606-8502, Japan}
\affiliation{ Kavli Institute for the Physics and Mathematics of the Universe (WPI), The University of Tokyo Institutes for Advanced Study, The University of Tokyo, Kashiwa, Chiba 277-8583, Japan}

\begin{abstract}
We study a theory of massive tensor gravitons which predicts blue-tilted and largely amplified primordial gravitational waves. After inflation, while their mass is significant until it diminishes to a small value, gravitons are diluted as non-relativistic matter and hence their amplitude can be substantially amplified compared to the massless gravitons which decay as radiation. We show that such gravitational waves can be detected by interferometer experiments, even if their signal is not observed on the CMB scales. 
\end{abstract}
\maketitle


\section{I. Introduction}

Cosmic inflation became a standard paradigm in primordial cosmology, while
it is still the subject of intensive researches for its unknown nature. 
A major prediction of the inflation theory is the production of primordial gravitational waves (GWs) which is scale-invariant and whose amplitude
is proportional to the inflationary Hubble scale.
Thus, by measuring the amplitude, we can reveal the energy scale of inflation.
A number of different experiments such as Planck~\cite{Akrami:2018odb}, 
SKA~\cite{Janssen:2014dka}, 
LISA~\cite{AmaroSeoane:2012je},  
Advanced-LIGO (A-LIGO)~ \cite{TheLIGOScientific:2016dpb}
 and 
DECIGO~\cite{Seto:2001qf,Kawamura:2011zz} put bounds on or aim to detect it.
Nevertheless, it should be stressed that even if inflation occurred, the primordial GWs may be different from the conventional prediction based on general relativity. 
Among various possibilities to generalize the gravity theory, 
massive gravity attracts conspicuous attention 
and has been applied to the study on the primordial
GWs~\cite{Dubovsky:2009xk, Gumrukcuoglu:2012wt,Fasiello:2015csa, Kuroyanagi:2017kfx}.

The study of massive gravity stemmed from one of fundamental questions in classical field theory, ``Can a spin-$2$ field have a non-vanishing mass or not?'' This led Fierz and Pauli in 1939~\cite{Fierz:1939ix} to find a unique Lorentz-invariant mass term for a linearized spin-$2$ field, for which a nonlinear completion was found in 2010~\cite{deRham:2010ik,deRham:2010kj}.
Another motivation is the accelerated expansion of the universe today: a graviton mass term may lead to acceleration without a need for dark energy. From this point of view, the assumption of Lorentz-invariance does not seem to have a firm justification since the graviton mass as an alternative to dark energy is supposed to be of the cosmological scale today and the expansion of the universe anyway breaks the Lorentz-invariance at the cosmological scale.

Once the assumption of Lorentz-invariance is relaxed at the cosmological scale, new possibilities open up~\cite{ArkaniHamed:2003uy,Rubakov:2004eb,Dubovsky:2004sg,Blas:2009my,Comelli:2013txa,Langlois:2014jba}. In particular, a massive graviton forms a representation of the three-dimensional rotation group instead of four-dimensional Lorentz group, and therefore the number of physical degrees freedom in the gravity sector does not have to be five. The minimal theory of massive gravity (MTMG) introduced in \cite{DeFelice:2015hla,DeFelice:2015moy} is one of such possibilities and propagates only two physical degrees of freedom in the gravity sector, allowing for self-accelerating, homogeneous and isotropic cosmological solutions without pathologies such as strong coupling and ghosts, that are usually unavoidable in Lorentz-invariant massive gravity~\cite{DeFelice:2012mx}. The recently developed positivity bounds that significantly shrink the viable parameter space of the Lorentz-invariant massive gravity theory \cite{Cheung:2016yqr,Bonifacio:2016wcb,Bellazzini:2017fep,deRham:2017xox} also do not apply to those Lorentz-violating theories, including MTMG, since those bounds rely on Lorentz invariance at all scales. Moreover, because of the absence of extra degrees of freedom, MTMG completely evades the so called Higuchi bound, which states that the mass of a Lorentz-invariant massive graviton should be greater than the Hubble expansion rate up to a factor of order unity in order to avoid turning extra degrees of freedom into ghosts in cosmological backgrounds~\cite{Higuchi:1986py}. From the viewpoint of effective field theories, it is plausible to expect that there should be other Lorentz-violating massive gravity theories with similar properties and MTMG is just one concrete example of such theories. 
\footnote{
A generalization of solid inflation~\cite{Gruzinov:2004ty, Endlich:2012pz} dubbed supersolid inflation~\cite{Nicolis:2013lma}  is classified
into these theories and the primordial GWs in supersolid inflation are studied in \cite{Cannone:2014uqa, Bartolo:2015qvr, Ricciardone:2016lym, Ricciardone:2017kre}.}
As we shall see in the rest of the present paper, those properties stated here open up a new observational window to GWs produced in the early universe.

\section{II. Setup}

Our quadratic Lagrangian density for the tensor graviton $h_{ij}(\tau,\vec{x})$
is given by
\begin{align}
\mathcal{L}_h^{(2)}=&\frac{a^2M_{\rm Pl}^2}{8}
\big[h_{ij}' h_{ij}'
-\partial_lh_{ij}\partial_lh_{ij}-a^2
\mu^2h_{ij}h_{ij}\big],
\label{Lh}
\end{align}
where prime denotes a derivative with respect to
the conformal time $\tau$,
$M_{\rm Pl}$ is the reduced Planck mass,
$a(\tau)$ is the scale factor and $\mu(\tau)$
is the mass of the tensor graviton which depends on time.
The time dependence of $\mu(\tau)$ may originate in a
dynamics of other fields (e.g. a homogeneous scalar field $\mu(\varphi(\tau))$),
while we do not discuss any concrete model in this letter.

The tensor gravitons can be decomposed as
\begin{equation}
h_{ij}=\frac{2}{aM_{\rm Pl}}\sum_{\lambda=+,\times}
\int \frac{\mathrm{d}^3 k}{(2\pi)^3} e^{i\bm{k\cdot x}}e_{ij}^\lambda  \left[v^\lambda_{k}(\tau)\hat{a}^\lambda_{\bm k}+{\rm h.c.} \right],
\end{equation}
where $e_{ij}^\lambda(\hat{\bm k})$ is the polarization tensor and $\hat{a}_{\bm k}/\hat{a}^\dag_{\bm k}$ are creation/annihilation operators satisfying the commutation relation,
$[\hat{a}^\lambda_{\bm k}, \hat{a}^{\dag\sigma}_{\bm p}]=(2\pi)^3\delta^{\lambda\sigma} \delta(\bm k-\bm p) $.
From the above action and the decomposition, one finds that the equation of motion (EoM) for the mode function $v_k^\lambda(\tau)$ is
\begin{equation}
v_k''+\left[k^2 +a^2\mu^2-\frac{a''}{a}\right]v_k=0,
\label{v EoM}
\end{equation}
where we have suppressed the polarization label $\lambda$ because the EoM does not depend on it.

To solve the above EoM, we need to specify $a(\tau), \mu(\tau)$ and the initial condition for $v_k(\tau)$.
For simplicity, we assume the de Sitter expansion $a \propto \tau^{-1}$ during inflation as well as instantaneous reheating followed by the radiation dominated era $a\propto\tau$. Then the scale factor is written as 
\begin{equation}
a(\tau)=\begin{cases}
-1/(H_{\inf}\tau) & (\tau<-\tau_r) \\
a_r\tau/\tau_r & (\tau>\tau_r) \\
\end{cases},
\end{equation}
where $H_{\inf}$ is the Hubble expansion rate during inflation and
$a_r$ is the scale factor at the reheating time $\tau_r= (a_r H_{\inf})^{-1}$.
Note that the conformal time $\tau$ jumps from $-1/(aH_{\inf})$ into $1/(aH_{\inf})$ at reheating in this treatment for $a$ and $\mathrm{d} a/\mathrm{d} \tau$ to be continuous. We further assume a simple step-function behavior of the graviton mass,
\begin{align}
\mu(\tau)&=
\begin{cases} 
m & (\tau<\tau_m) \\
0 & (\tau>\tau_m) \\
\end{cases},
\label{m evolution}
\end{align}
where $\tau_m$ is a certain time during radiation dominated era.%
\footnote{In principle, gravitons can remain significantly massive for 
redshift $z\gtrsim 10^{-2}$, because only the recently detected
binary neutron star merger (i.e. GW170817 and GRB170817A) occurred at $z\simeq 10^{-2}$ puts a direct bound on the propagation speed of gravitons~\cite{Monitor:2017mdv}.
However, we conservatively assume that the graviton mass vanishes before
the matter-radiation equality in this letter.}
Finally, we set the initial condition for the mode function to be that for the Bunch-Davies vacuum during inflation,
\begin{equation}
\lim_{k\tau\to -\infty}v_k(\tau) = \frac{1}{\sqrt{2 k}}e^{-i k\tau}.
\label{BD condition}
\end{equation}
These conditions suffice to obtain the evolution of the massive tensor gravitons in our setup.

\section{III. Evolution}

In this section, we shall study the time evolution of the tensor
massive gravitons and calculate their dimensionless power spectrum,
\begin{equation}
\mathcal{P}_h(\tau, k) = \frac{4k^3|v_k(\tau)|^2}{\pi^2 M_{\rm Pl}^2 a^2(\tau)  },
\label{Ph def}
\end{equation}
where the contributions from the two polarization have been summed.
The mode function of the tensor gravitons with a long wave length 
changes its behavior twice; namely
at the end of inflation and when their mass vanishes.
Therefore we have the following three phases, (i) inflation phase $\tau<\tau_r$,
(ii) mass dominant phase
$\tau_r<\tau<\tau_m$ and 
(iii) massless phase $\tau_m<\tau$.
We shall discuss these phases in order.

\vspace{2mm}
{\bf (i) Inflation phase}:
Solving the EoM \eqref{v EoM} in the de Sitter universe with the initial condition \eqref{BD condition}, one finds the solution of the mode function during inflation as
\begin{equation}
 v_k^{({\rm i})}(\tau)= \frac{\sqrt{-\pi \tau}}{2} H_\nu^{(1)}(- k \tau)\,, \quad
  \nu\equiv \sqrt{\frac{9}{4}-\frac{m^2}{H^2_{\inf}}}\,,
\label{inf sol}
\end{equation}
where $H_\nu^{(1)}(z)$ is Hankel function of the first kind.
In the super-horizon limit $-k\tau\to 0$, it asymptotes $v_k^{({\rm i})}\propto \tau^{\frac{1}{2}-\nu} k^{-\nu}$.
Thus, massive tensor gravitons produce blue-tilted tensor power spectrum $\mathcal{P}_h \propto (\tau k)^{3-2\nu}$ which
decreases on super-horizon scales due to the graviton mass
during inflation, $\nu<3/2$. The usual scale-invariant spectrum 
is restored in the massless limit $\nu\to 3/2$.
This result can be understood as an analogy to the fluctuation of a massive scalar field.

\vspace{2mm}
{\bf (ii) Mass dominant phase}:
Next, let us discuss the evolution of the massive tensor gravitons after inflation. 
The $a''/a$ term in \eqref{v EoM} vanishes  
during the radiation dominated era, and almost all the modes which exited the horizon during inflation satisfy
$k\ll am$  for $m/H_{\inf}=\mathcal{O}(1)$.
Hence the EoM in this phase reads 
\begin{equation}
\partial_\tau^2 v_k^{({\rm ii})}+a^2m^2 v_k^{({\rm ii})}\simeq 0.
\end{equation}
We expect that these modes 
behave as non-relativistic matter. 
By using the junction condition,
\begin{equation}
v_k^{({\rm i})}(-\tau_r)=v_k^{\rm (ii)}(\tau_r),
\quad
\partial_\tau v_k^{({\rm i})}(-\tau_r)=\partial_\tau v_k^{\rm (ii)}(\tau_r).
\label{junction condition}
\end{equation}
one finds that the mode function shows
a damped oscillation at sufficiently late times $m\tau^2\gg H_{\inf}\tau_r^2$,
\begin{equation}
v_k^{\rm (ii)}\simeq \frac{2}{\sqrt{\pi am}}\left[
C_1 \cos\left(\frac{am\tau}{2}-\frac{\pi}{8}\right)
+C_2 \sin\left(\frac{am\tau}{2}+\frac{\pi}{8}\right)
\right].
\label{vmass}
\end{equation}
with the integration constants, 
$C_{1,2} \simeq -i\sqrt{\pi} 2^{-\frac{7}{2}+\nu}$ $(k\tau_r)^{-\nu}\Gamma(\nu)
\left[ \frac{2m}{H_{\inf}}\, J_{\mp \frac{3}{4}}\left(\frac{m}{2H_{\inf}}\right)\pm (1-2\nu) \,J_{\pm\frac{1}{4}}\left(\frac{m}{2H_{\inf}}\right)\right]$.
Here $J_\mu (z)$ denotes the Bessel function of the first kind.

 \eqref{vmass} implies that $v_k\propto a^{-1/2}$ in this phase
and hence the graviton energy density evolves like non-relativistic matter 
$m^2 h_k^2 \propto a^{-2}v_k^2\propto a^{-3}$ as expected.
Compared to the massless graviton whose energy density
decays as $a^{-4}$, the decay of the massive tensor graviton
is slower and thus its final amplitude will be relatively amplified.

\vspace{2mm}
{\bf (iii) Massless phase}:
At $\tau=\tau_m$, the graviton mass disappears and gravitons restore their normal behavior as radiation whose energy density decays as $a^{-4}$.
By using the junction conditions same as  \eqref{junction condition}
at $\tau=\tau_m$ between $v_k^{({\rm ii})}$ and $v_k^{({\rm iii})}$, we obtain the mode function as
\begin{equation}
v_k^{({\rm iii})} (\tau)=\frac{2}{k}\sqrt{\frac{m\tau_m}{\pi H_{\inf} \tau_r^2}} 
\Big[D_1 \cos(k\tau)+D_2\sin(k\tau)\Big],
\end{equation}
with $D_1\simeq-\sin(k\tau_m)[C_2 \cos(\Lambda+\pi/8)-C_1\sin(\Lambda-\pi/8)],$
$D_2\simeq \cos(k\tau_m)[C_2 \cos(\Lambda+\pi/8)-C_1\sin(\Lambda-\pi/8)]$
and $\Lambda\equiv m\tau_m^2/(2H_{\inf}\tau_r^2)$.
Here we used $\Lambda\gg k\tau_m$. 
In this phase, the mode function significantly grows and then starts oscillating with a constant amplitude when it re-enters the horizon. 
In the case with $m/H_{\inf}=\mathcal{O}(1)$,
the squared amplitude is roughly given by
\begin{equation}
2k|v_k^{({\rm iii})}|^2 \sim \frac{\tau_m}{\tau_r} (k\tau_r)^{-2\nu-1}.
\end{equation}
Compared to the usual massless graviton with $\nu=3/2$ and
$\tau_m\to \tau_r$, the power spectrum of our massive graviton  is amplified by the two factors: the first factor $\tau_m/\tau_r$ represents the duration of the epoch where the decay of the graviton energy density is slower by $a\propto \tau$; the second factor $(k\tau_r)^{-2\nu-1}$ for $\nu<3/2$ represents the damping effect during inflation which leads to a blue-tilted spectrum.
Therefore our final power spectrum of the primordial GWs for $\tau>\tau_m$ can be evaluated as
\begin{equation}
\mathcal{P}_h^{\rm massive}(\tau) \sim 
\frac{\tau_m}{\tau_r}(k\tau_r)^{3-2\nu}\,
\mathcal{P}_h^{\rm massless}(\tau),
\label{Power relation}
\end{equation}
where $\mathcal{P}_h^{\rm massless}$ denotes the usual power spectrum of the massless tensor modes from inflation.

\section{IV. Results}

Now we study the parameter region in which the primordial GWs
generated in our scenario satisfy the current constraints 
and can be observed by upcoming experiments. To this end, we consider
the energy fraction of the GWs per logarithmic interval of the wave number $k$ at the present time,
$\Omega_{\rm GW,0}(k)\equiv \rho_{\rm tot}^{-1}\,\mathrm{d}\rho_{\rm GW}/\mathrm{d} \ln k $.
From  \eqref{Power relation} and using $\Omega_{\rm GW,0}^{\rm massless}(k)\sim 10^{-15}H_{14}^2$ for the modes which entered the horizon during the radiation dominated era, one finds 
\begin{align}
\Omega_{\rm GW,0}^{\rm massive}(k)
&\sim \frac{\tau_m}{\tau_r}(k\tau_r)^{3-2\nu}
\Omega_{\rm GW,0}^{\rm massless}(k),
\notag\\
&\approx 10^{-15}\,\frac{\tau_m}{\tau_r}H_{14}^{\nu+\frac{1}{2}} f_8^{3-2\nu}
,
\label{OGW estimate}
\end{align}
where $H_{14}\equiv H_{\inf}/(10^{14}{\rm GeV})$ and $f_8\equiv f/(2\times10^8{\rm Hz})$.
Here the GW frequency $f\equiv k/2\pi$  is assumed to be lower than the inflationary UV cutoff, 
$f_{\rm UV}\simeq  a_r H_{\inf}/(2\pi)\approx2 \times 10^8H_{14}^{1/2}\,{\rm Hz}$ and higher than the scale corresponding to the matter-radiation equality,  $f_{\rm eq}\simeq\ 3\times 10^{-17}$Hz.

The BBN bound $\Omega_{\rm GW,0}<10^{-5}$ and the CMB bound
$\Omega_{\rm GW,0}(f\sim 2\times 10^{-17}\,{\rm Hz})<10^{-15}$ are recast as
\begin{align}
\frac{\tau_m}{\tau_r}&\lesssim 10^{10} H_{14}^{-2}, \quad  (\rm BBN)
\label{BBN bound}\\
\nu &\lesssim \frac{75-\log_{10}(H_{14}^{1/2}\tau_m/\tau_r)}{50+\log_{10}(H_{14})}.
\quad  (\rm CMB)
\label{CMB bound}
\end{align}
The largest GWs can be produced when these two conditions are saturated for given $H_{\inf}$. In that case, the graviton mass during inflation is given by
\begin{equation}
\frac{m^2}{H_{\inf}^2}\bigg|_{\rm max\, GW}
\simeq \frac{9}{4}-\left(\frac{65+\frac{3}{2}\log_{10}H_{14}}{50+\log_{10}H_{14}}\right)^2.
\end{equation}
%

%
\begin{figure}[tbp]
  \includegraphics[width=90mm]{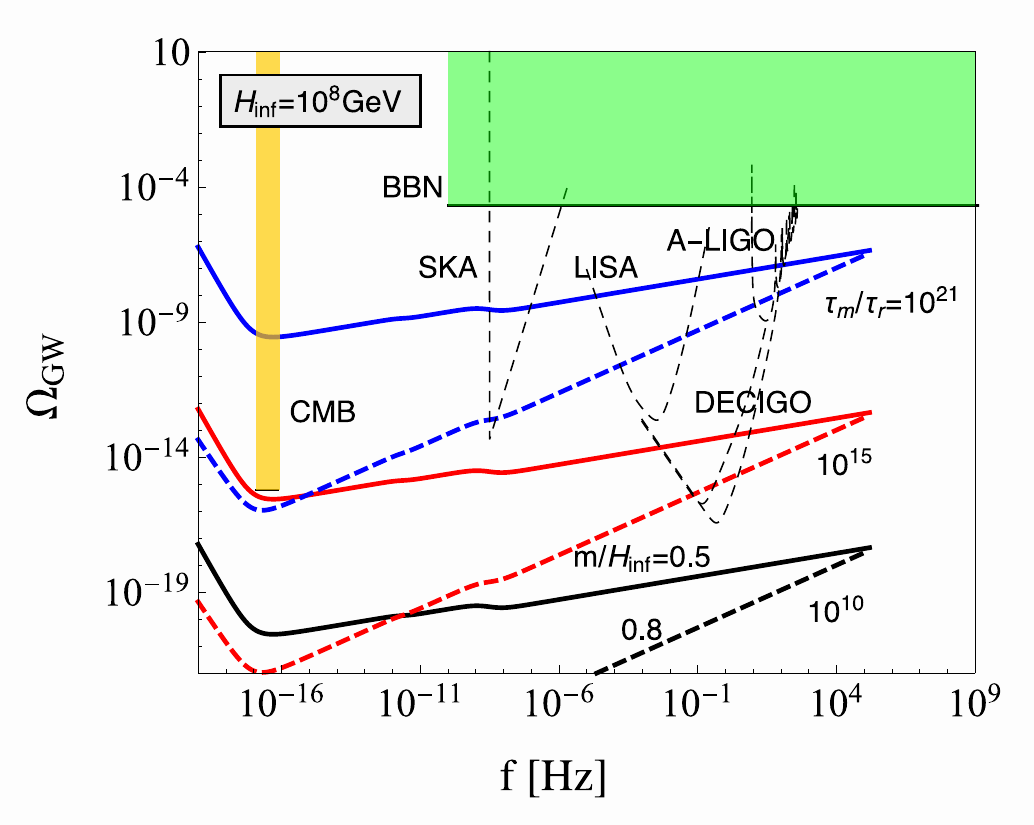}
  \caption
 {We plot $\Omega_{\rm GW}$ for $H_{\rm inf}=10^8{\rm GeV}$ and  $\tau_m/\tau_r=10^{10}$ (black), $10^{15}$ (red) and $10^{21}$ (blue) as thick lines. 
The graviton mass is $m=0.5H_{\inf}$ (solid) and $m=0.8H_{\inf}$ (dashed). The shaded regions are excluded by the BBN (green) and CMB (yellow) constraints. The sensitivity curves of SKA, LISA, A-LIGO and DECIGO (with the original and upgraded sensitivity curves \cite{Kuroyanagi:2014qza}) are also shown
as thin dashed lines.}
 \label{fig_spectrum}
\end{figure}
%

It should be noted that gravitons are still massive at BBN if
\begin{equation}
\frac{\tau_m}{\tau_r}\gtrsim \frac{\tau_{\rm BBN}}{\tau_r}=\sqrt{\frac{H_{\inf}}{H_{\rm BBN}}}\approx 10^{17}H_{14}^{\frac{1}{2}}.
\end{equation}
In this case, the BBN bound \eqref{BBN bound} should be relaxed, because
gravitons do not contribute to relativistic degrees of freedom during BBN.
In this letter, however, we conservatively respect the original bound \eqref{BBN bound}.
Whereas, since we assumed the graviton mass vanishes at $\tau_m$ before the 
radiation-matter equality at $\tau_{\rm eq}$, the amplification factor 
$\tau_m/\tau_r$ cannot exceed,
\begin{equation}
\frac{\tau_m}{\tau_r}\le \frac{\tau_{\rm eq}}{\tau_r}=\sqrt{\frac{H_{\inf}}{H_{\rm eq}}}\approx 2\times 10^{25}H_{14}^{\frac{1}{2}}.
\label{equality limit}
\end{equation}

\begin{figure}[tbp]
  \includegraphics[width=80mm]{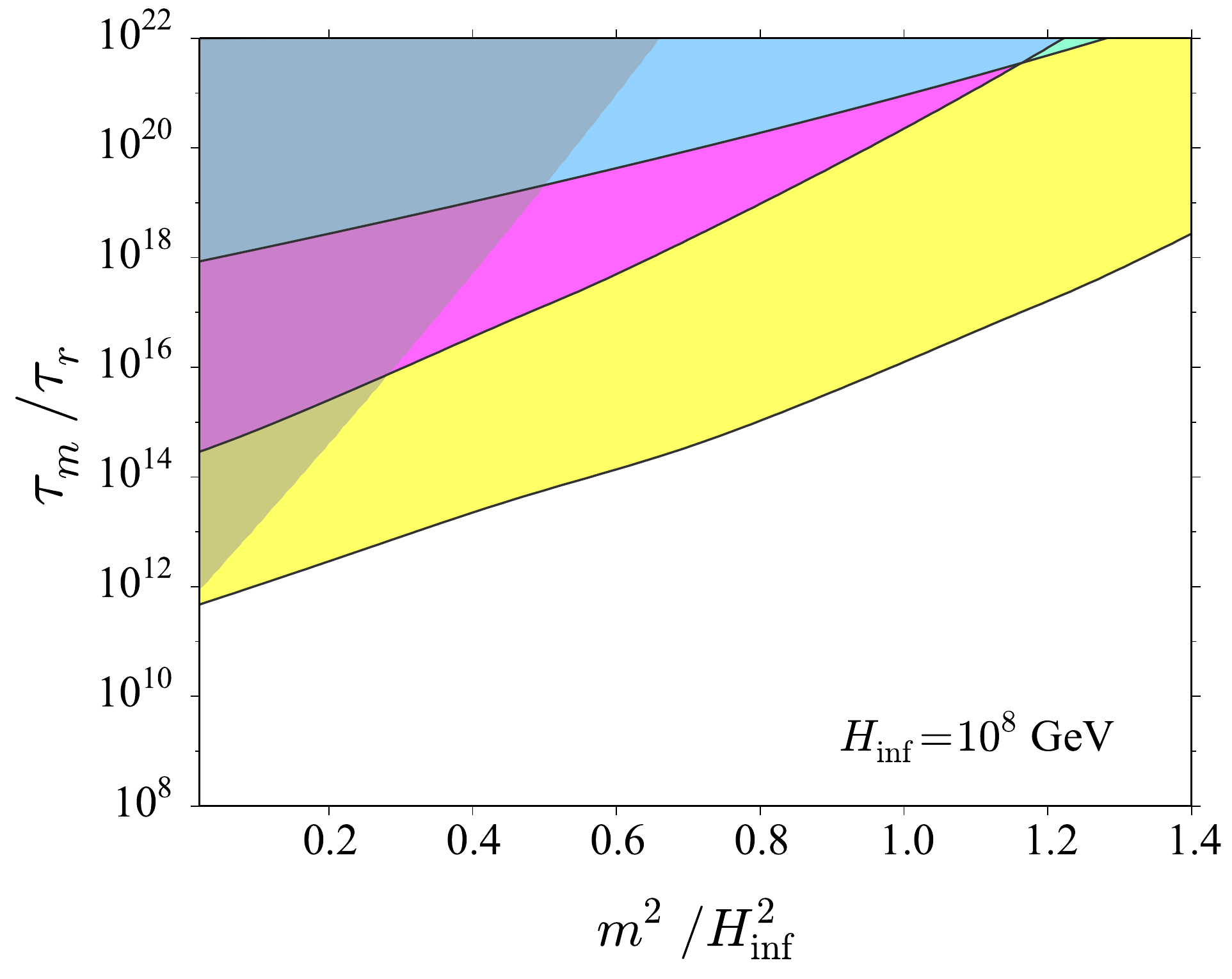}
  \caption
 {The GWs can be detected only by DECIGO (yellow), by LISA and DECIGO (pink),  by aLIGO and DECIGO (green) and by all of aLIGO, LISA and DECIGO (blue)
 in the coloured parameter regions. The grey shaded region is excluded by the
CMB observation. The Hubble scale is fixed at $H_{\inf}=10^8{\rm GeV}$ which
corresponds to $\Omega_{\rm GW} \sim 10^{-27}$ in the massless graviton case.}
 \label{fig_contour}
\end{figure}

In Figure.~\ref{fig_spectrum}, we show the accurate prediction for the primordial GWs by numerically solving the EoM and compare them with the sensitivity curves of the various experiments.
Although the inflationary Hubble scale is chosen as low as $H_{\inf}=10^8{\rm GeV}$,
the amplification during the mass dominant phase with $\tau_m/\tau_r\gg1$
makes the GWs large enough to be observed. Furthermore, the blue-tilted spectra made by the graviton mass naturally enable to avoid the CMB constraint
and to have detectable amplitudes on smaller scales at the same time.
It is interesting to note that all of SKA, LISA and aLIGO can detect
the GWs for $m=0.8H_{\inf}$ and $\tau_m/\tau_r=10^{21}$ (see the blue dashed line).
Except for the blue solid line which is excluded by the CMB constraint, all the lines satisfy other constrains including ones from the PTA~\cite{Lentati:2015qwp} and the CMB $\mu$ distortion~\cite{Ota:2014hha} which are not shown in Figure.~\ref{fig_spectrum}.
In Figure.~\ref{fig_contour}, the detectable parameter region of ($m$, $\tau_m/\tau_r$) is explored.
For sufficiently large $m$ and $\tau_m/\tau_r$, the detectability and the CMB constraint are compatible, while a too large
$\tau_m/\tau_r$ violates \eqref{equality limit} and requires to revisit the BBN bound.
%
%

\section{V. Discussion}

In this section, we discuss a possibility to further extend the model based on the effective field theory (EFT) approach. We also discuss the connection between the graviton mass and the dark energy based on the minimal theory of massive gravity (MTMG).

Although we have assumed the quadratic Lagrangian \eqref{Lh} so far, according to the general philosophy of the EFT approach, one should consider a non-trivial sound speed of graviton $c_T$ introduced as $-c_T^2\partial_lh_{ij}\partial_lh_{ij}$ in \eqref{Lh}.
Provided that $c_T<1$ is constant for $\tau<\tau_m$ and becomes unity at $\tau=\tau_m$ in the same way as the graviton mass, the varying $c_T$ leads to the following three modifications.
(i) The  UV cutoff frequency $f_{\rm UV}$ increases by $c_T^{-1}$.
(ii) The tensor power spectrum is amplified by $c^{-2\nu}_T$ for $k\ll a(\tau_m)m$.
(iii) For the modes with $a(\tau_m) m\lesssim k$, which are produced only if $c_T\lesssim \tau_r/\tau_m$, the tilt of the tensor power spectrum becomes bluer.
A detailed study on the cases with a non-trivial $c_T$ is left for future work.
Supersolid inflation~\cite{Nicolis:2013lma} based on EFT can also generate a highly blue-tilted GWs \cite{Ricciardone:2017kre}. 
However, the post-inflationary dynamics was not considered and the amplification mechanism was missed in the work.
It would be interesting to combine such models with our analysis.

In the present paper we have studied impacts of a class of massive gravity theories on GWs observations without specifying a concrete theory. This is a totally rational attitude from the viewpoint of EFTs. It is nonetheless interesting to discuss concrete examples. Here we thus consider one such example based on MTMG~\cite{DeFelice:2015hla,DeFelice:2015moy}. 
The FLRW cosmology in this theory has two branches of solutions, the self-accelerating branch and the normal branch. In the former branch the effective cosmological constant is
\begin{equation}
 \Lambda_{\rm eff} = \frac{m_{\rm g}^2}{2}X(c_1X^2+3c_2X+3c_3)\,,
\end{equation}
where $c_i (i = 1,2,3)$ are dimensionless constants in the gravity action,
$m_{\rm g}$ is a mass scale and $X$ is a constant satisfying $c_1 X^2 + 2c_2X + c_3 = 0$.
The graviton acquires a squared mass,
\begin{equation}
 \mu^2 = \frac{m_{\rm g}^2}{2} X\left[c_2X+c_3 + \frac{H}{H_f}(c_1X+c_2)\right]\,,
\end{equation}
where $H_f$ is the Hubble expansion rate of the fiducial metric that can be freely specified as a part of the definition of the model. 
One can promote the constants $c_i$ to functions of a scalar field,
$c_i = c_i (\phi).$ When $\phi$ is constant, $\Lambda_{\rm eff}$ and $\mu^2$ in the self-accelerating branch are given by the above
formulas. If $\phi$ starts with a constant and changes to another constant then $\Lambda_{\rm eff}$
and $\mu^2$ also exhibit a transition. By tuning $c_i (\phi)$ and $H_f$ , one can in principle realize such a model that $\mu^2 \gg |\Lambda_{\rm eff}|$ before the transition.

\section{VI. Conclusion}

In this letter, we have investigated the primordial GWs in
the theory of massive tensor gravitons \eqref{Lh}.
Contrary to the massless graviton case, the massive gravitons with
a mass comparable to the inflationary Hubble scale $m=\mathcal{O}(H_{\inf})$ generate a blue-tilted tensor spectrum during inflation. 
Moreover, while their mass is significant after inflation, the dilution of the energy density of the massive gravitons becomes slower $\rho_h^{\rm massive}\propto a^{-3}$ than the massless ones $\rho_h^{\rm massless} \propto a^{-4}$.
Thus, $\Omega_{\rm GW,0}$ in the massive case can be substantially amplified compared to the massless case. 
Consequently, we have obtained the blue-tilted and largely amplified primordial
GWs which are suitable for the detection by the interferometers and to avoid the CMB constraint at the same time.
We have derived the analytic expression for $\Omega_{\rm GW,0}$  \eqref{OGW estimate} and illustrated its detectability in Fig.~\ref{fig_spectrum}
and \ref{fig_contour}. We have found that it is even possible to generate
primordial GWs detectable for all of SKA, LISA and advanced-LIGO.
Our findings further motivate the theoretical works on massive gravitons and the experimental efforts to detect stochastic GWs.

\section{Acknowledgement}
\label{Acknowledgement}
 We would like to thank Claudia de Rham, Tsutomu Kobayashi,
Misao Sasaki, Takahiro Tanaka and Gianmassimo Tasinato for useful comments.
 In this work, TF is supported by Japan Society for the Promotion of Science (JSPS) Grants-in-Aid for Scientific Research (KAKENHI) No.~17J09103. 
 SK is supported by JSPS KAKENHI No. 17K14282 and Career Development Project for Researchers of Allied Universities.
The work of SMi was supported by JSPS KAKENHI No. 16K17709. 
The work of SMu was supported by JSPS KAKENHI No. 17H02890, No. 17H06359, and by WPI, MEXT, Japan.

\end{document}